# Methodology of selective metallic thin film ablation from susceptible polymer substrate using pulsed femtosecond laser

Chresten von der Heide [a], Maria Grein [b], Günter Bräuer [b], Andreas Dietzel [a]
[a] Technische Universität Braunschweig, Institute of Microtechnology, Braunschweig, 38124, Germany
[b] Technische Universität Braunschweig, Institute for Surface Technology, Braunschweig, 38108, Germany
c.von-der-heide@tu-braunschweig.de, maria.grein@ist-extern.fraunhofer.de

**Abstract**

Electronic devices are progressively fabricated on flexible substrates allowing new fields of applications. In the future flexible substrates will be found in many life science and medical products as they adapt to given shapes and integrate well into all types of soft materials. A maskless and very flexible structuring process is offered by ablation using ultra-short pulse laser irradiation. Usually, certain areas of a functional thin film coating are locally removed to yield the needed device structures. Micro laser patterning quality is not only influenced by the beam properties (beam profile, fluence) but also by pulse overlap, scan repetitions and many other factors such as substrate and coating material. This makes process parameter optimization a challenging task. In this paper, we present a systematic approach to efficiently find suitable parameters for laser-induced ablation of thin films on susceptible polymer substrates. As an example, we use a sputtered NiCr coating with a thickness of 100 nm on a polyimide film made by spin coating of PI-2611 precursor (@2000 rpm → ca. 8 µm, HD microsystems). Irradiation is conducted using a fs-laser with infrared wavelength of 1030 nm and a pulse length of 212 fs. The energy per pulse is varied in the range of 0,29 µJ to 4,83 µJ, yielding fluence values between 70 mJ/cm$^2$ and 1,11 J/cm$^2$. Ablation threshold fluence for a 100 nm layer of NiCr was found to be $\phi_{th}(N_P: 1) = 0,50$ J/cm$^2$ and the incubation factor was evaluated to be $\xi = 0,53$. A clear distinction must be made between the material ablation at the surface of a bulk material and the selective removal of a thin film. In the second case effects such as flaking occur in practice and influence the ablation characteristic. We then investigate different varieties of thin film removal including dot-, line- and areal-ablation. The methodology is presented using a practical example, but can be applied to selective ablation of a wide range of thin film systems.

## 1 Introduction

Nowadays, metallic thin films and coatings are an essential part of many advanced systems but also everyday devices [1, p. 1157]. In the past decade a technological revolution was observed. Electronic devices, in former times solely based on solid printed circuit boards, are being altered such that their functional structures can be fabricated on flexible substrates [2]. This allows for new fields of applications e.g. live science or medical products; as flexible systems can be conformally applied and integrated into all kinds of soft materials. Flexible substrates are often made from polymers. Most polymer foils are available in transparent or opaque form and with a wide range of mechanical properties. Also, they show good dielectric properties and are relatively cheap to produce. For the fabrication of electronic devices mainly thermal and chemical robustness of the chosen polymer substrate have to be considered, because many microelectronic production processes are based on lithography and chemical etching [3, p. 1123]. The use of etchants, solvents or heat to cure fotosensitive resins can reduce the selection of suitable substrate materials. As an alternative, the ablation by ultra-short pulse laser irradiation provides a maskless and very flexible process for structuring thin films on flexible polymeric substrates [4, p. 7575]. In general, highly energetic ultrashort pulses can be used to ablate material with marginal thermal input to the substrate, whereby the heat-affected volume in the vicinity of the ablation area is minimized [5, p. 1706] [1, p. 1157].

The quality of micro laser patterning can be controlled via pulse overlap, fluence and scan repetitions [1, p. 1157]. For flexible electronic systems usually certain areas of a conductive thin film coating are removed to yield the needed functional (sensor-) structures. Thus, complete galvanic insulation between separated structures has to be accomplished [6]. Precise movement of the laser beam along vertically stacked trajectory patterns allows areal ablation with pulsed laser systems. Despite the fact that single pulse ablation thresholds can be precisely determined with known techniques, the influence of scan parameters for line- and areal ablation is less investigated. Instead, trial-and-error parameter grids are mostly used in the field. Of course, these can also lead to a useful set of parameters whereby the ablation results are often evaluated according to rather subjective quality standards. However, the avoidance of substrate damage is crucial and is therefore regarded as the main optimization parameter in the context of this paper. For polyimide as a substrate, it can be shown that even minor substrate damage can be detected by purely optical measuring methods. The known methodology of ablation threshold determination was developed for rather bulky materials with the goal of cutting down into a substrate to a certain depth [7, p. 239]. However, differences between thin film ablation and bulk material abrasion through laser induced evaporation of (bulk-) material cannot be ruled out. Thus, methods and criteria allowing predictions for selective ablation results are unclear or missing.

Ideal selective ablation is achieved, when the thin film coating is removed without residues while the substrate stays untouched. This is especially challenging for metal coatings on susceptible polymer substrates. Because of this difficulty, authors in the past have considered small substrate damage (e.g. 1 µm abrasion from a 50 µm polymer foil after ablating a 150 nm metallic thin film) inevitable to achieve proper electrical insulation [8]. As this is justified for some applications, it might limit the use of laser ablation processes when areal removal on transparent substrates should be accomplished as these lose their transparency and become foggy even after slight surface damaged. Also mechanical problems can arise due to the weakened substrates. To evade these limitations we will present a systematic methodology for laser parameter identification and laser scan strategies which allows to achieve selective ablation which can no longer be distinguished from the ideal selective case by the optical methods used. Our parameter selection approach is based on finding the sample/material specific ablation fluence threshold $\phi_{th}$ via multiple single spot D²-experiments. Two criteria will be introduced to predict ablation results such that the parameter search can be narrowed down beforehand.

## 1.1 Laser and inspection equipment

All laser based experiments are carried out using a pulsed Yb:KGW (Ytterbium-doped Potassium-Gadolinium Tungstate) solid-state laser system (microSTRUCT C, 3DMicromac), with a pulsed laser source ($\nu = 212$ fs) emitting at a primary wavelength of $\lambda = 1030$ nm in linear, horizontal polarization. Pulse frequency is 600 kHz for most ablation experiments. Beam quality is specified with $M^2 < 1,1$. The laser beam has a diameter of of $\emptyset_{Beam} = 4,7$ mm and can further be expanded using a Galilei lens setup. High speed laser beam positioning on the specimen is provided by a galvanometer-scanner (Scanlab intelliSCAN 14), while focusing is accomplished by a f-theta telecentric lens (Linos F-Theta Ronar) with an effective focal length of 100,1 mm and a apodisation factor of 1,83. The laser source is operated at a reduced power of 3W providing pulse energies between 0,28 and 4,85 µJ. All testing is carried out under normal ambient conditions at room temperature.

Analysis of single spot experiments as well as inspection and evaluation of ablation quality is conducted with a confocal laser scanning microscope (CLSM, Keyence, VK-X260K). Its microscope objective provides a magnification of 150x at a numerical aperture of 0,95 which enables morphological examination of the surface.

## 1.2 Sample preparation

We use polyimide substrates on rigid class carrier wafers for better handling. Polyimide foils are prepared by spin coating a liquid precursor (PI-2610/11, HD microsystems) onto the carrier as described by [9]. An adjacent thermal treatment is used for curing. Prior to sputter deposition, the foil surface is treated by inverse sputter etching to improve adhesion of the coating. A layer of 100 nm of Nickel-Chromium (target composition: 80 wt% Nickel and 20 wt% Chromium) as piezoresistive metal thin film alloy commonly used for strain gauge applications [10, p. 2], is deposited on the substrate.

## 2 Calculation of fluence distributions

An areal ablation process with a pulsed laser source is accomplished by moving the beam in a scan pattern over the surface of the specimen. A combination of scanner speed v, pulse frequency f and spot radius $\omega_0$ determines the resulting pulse overlap parameter $\varphi$ (complete overlap at $\varphi = 0$ and no overlap for $\varphi > 1$).

$$\varphi = \frac{l_P}{2 \cdot \omega_0} \qquad l_p = \frac{v}{f} \qquad (2\text{-}1)$$

In literature, pulse overlaps of at least 90% ($\varphi = 0,1$) are commonly used to reduce edge line roughness. In this paper, we also work with much smaller overlaps as these will show to be advantageous when low average laser energy per ablation area is required. The real laser spot size can either be measured (by knife-edge method), or calculated from the specified beam width $\emptyset_{Beam}$ leaving the laser source. Our laser system provides a beam diameter of $\emptyset_{Beam} = 4,7$ mm measured with a beam profiler (Ophir Spiricon SP928) using the D4σ-standard. The collimated beam goes into the scanner unit and exits through a telecentric f-theta objective with an effective focal length (EFL) of 100,1 mm and an apodizationfactor of 1,83. This leads to a spot size $\emptyset_0$ of:

$$\emptyset_{0,calc} = 2 \cdot \omega_0 = \frac{1,83 \cdot \lambda \cdot \text{EFL}}{\emptyset_{Beam}} = 40,1 \text{ µm} \qquad (2\text{-}2)$$

By measuring the laserpower P with a thermal detector, the resulting pulse energy $E_p$ and maximum spot fluence in the beam center $\phi_0$ can be calculated as.

$$\phi_0 = \frac{2 \cdot P}{\pi \cdot \omega_0^2 \cdot f} = \frac{2 \cdot E_p}{\pi \cdot \omega_0^2} \qquad (2\text{-}3)$$

In our setup, the fluence profile along the radius r follows a spatial Gaussian distribution.

$$\phi(r) = \phi_0 \cdot e^{-2 \cdot \left(\frac{r}{\omega_0}\right)^2} \qquad (2\text{-}4)$$

Definitions such as FWHM, 4Dσ or $1/e^2$ coexist to specify the laser spotsize $\emptyset_0 = 2 \cdot \omega_0$. Here we use the latter two definitions which state that the border of the laserspot along the radius r in lateral direction is reached when the spot fluence has dropped to $\phi_0/e^2$. In the following $\phi_0$ will simply be called „fluence". In Fig. 1, a gaussian fluence distribution is sketched.

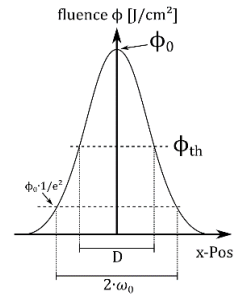

Fig. 1: Gaussian fluence profile

After scanning overlapping pulses along a straight line in X-direction, the resulting ablated cut width D can be thinner than the spotsize $\emptyset_0$ and depends on the specific material properties of the substrate.

In overlapping regions of two or more subsequent spots the fluence accumulates. For single laser lines the accumulated fluence profile can be gained by summing up the fluence distributions of individual pulses along its trajectory. The calculation of accumulated fluence profiles for areal ablation is more challenging, as the chosen ablation pattern has to be taken into account. We choose to work with a line pitch $l_{P,y}$ (Y-direction) of the stacked horizontal laser trajectories equal to the pulse spacing $l_{P,x}$ in the line scan direction X (see Fig. 2). This ablation pattern ($l_P = l_{P,x} = l_{P,y}$) will subsequently be called "#-pattern". It leads to an isotropic energy distribution.

For illustration purposes, "uncovered areas" that are not irradiated (when $\omega_0$ is interpreted as setting a sharp irradiation boundary) are marked in Fig. 2. In all our experiments the pulse overlap is chosen, such that full coverage of the ablation zone is achieved $0 < \varphi < 1$ in order to avoid these uncovered regions.

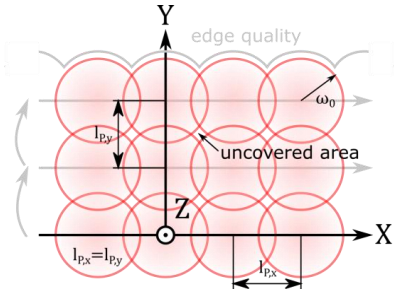

Fig. 2: Pulse placement for an isotropic energy distribution. The red circles indicate the spot size of single pulses with their $\omega_0$ boundaries. Grey lines indicate the order in which ablation pattern is progressing (starting at the abscissa moving up, left to right).

## 2.1 Accumulated fluence profile (AFP) as an irradiation model

Eichstädt et al. have introduced an accumulated fluence profile (AFP) which is valid for laser engraved lines as well as spatially extended areas [11, p. 82]. Thus, the fluence term includes x,y-coordinates shifted by multiples ($n_x$, $n_y$) of the distance between pulses ($l_P = l_{P,x} = l_{P,y}$):

$$\phi(x, y, n_x, n_y) = \phi_0 \cdot e^{-2 \cdot \frac{\left((x - l_P \cdot n_x)^2 + (y - l_P \cdot n_y)^2\right)}{\omega_0^2}} \quad (2\text{-}5)$$

$$n_x \in [0, \ldots, N_x], n_y \in [0, \ldots, N_y]$$

By summation of all successive pulse fluences the accumulated fluence profile $\Gamma(x, y)$ can be calculated as

$$\Gamma(x, y) = \sum_{n_X} \sum_{n_Y} \phi(x, y, n_x, n_y) \quad (2\text{-}6)$$

A line profile is obtained when one coordinate is not varied. The highest value in the set of data is considered as accumulated peak fluence $\Gamma_{AP}$ while the lowest value is called the accumulated valley fluence $\Gamma_{AV}$. We believe that $\Gamma_{AV}$ is of particular interest for the optimization of ablation processes. In addition, the homogeneity of the distributed laser energy must be considered. As shown in Fig. 3 the accumulated fluence profile of a laser line varies along its trajectory in X-direction. To avoid substrate damage, local high fluence maxima should be avoided, in particular if they exceed the substrate threshold fluence. To account for the fluence variance the overshoot variable $\theta_P$ is established, which relates $\Gamma_{AV}$ to $\Gamma_{AV}$:

$$\theta_P = \frac{\Gamma_{AP} - \Gamma_{AV}}{\Gamma_{AV}} \, [\%] \quad (2\text{-}7)$$

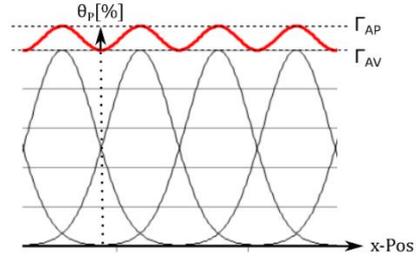

Fig. 3: Accumulated fluence profile (red) obtained from a line of laser spot intensities (grey).

Ahmmed et al. considered a fluence profile constant or flat if the overshoot value is smaller than 1 %, which can easily be achieved with large pulse overlaps [12, p. 259]. This is illustrated in Fig. 4, which shows different line accumulated fluence profiles for increasing $\varphi$. Fig. 4a represents the case $\varphi \geq 1$ where there is no overlap at all (spot size is specified by $\phi_0/e^2$ border). In this case, $\Gamma_{AP}$ equals $\phi_0$ but $\theta_P$ is large. Overlap can be increased to a level at which $\theta_P$ decreases but is still rather large while $\Gamma(x)$ practically not yet exceeds $\phi_0$ (Fig. 4b). With even further decrease of $\varphi$ (Fig. 4c) a point can be reached where $\Gamma_{AP}$ exceeds $\phi_0$, by a non-negligible amount. This point can be marked by the condition $\Gamma_{AV} = \phi_0$. Beyond that, the accumulated fluence level can raise far beyond $\phi_0$ (Fig. 4d) and eventually shows a very homogeneous fluence distribution, characterized by a very low $\theta_P$.

The fluence distribution in the case of 2D areal ablation patterns is illustrated in Fig. 5. Note that values for the accumulated fluence profile differ between line (1D)- and areal (2D)-ablation for the same value of $\varphi$. Assuming a standard areal ablation pattern in which laser lines (written in X-direction) are stacked above each other (in Y-direction), the accumulated fluence profile sums up by the contribution of successive pulses along the laser line as well as by pulses from vertically neighboring lines. Thus, $\Gamma$ will always be higher for 2D accumulated profiles than for 1D accumulated line profiles assuming the same value of $\varphi$. A homogeneously spread accumulated fluence profile (small overshoot variable $\theta_P$) is required to determine well-defined process parameters allowing a controlled areal ablation.

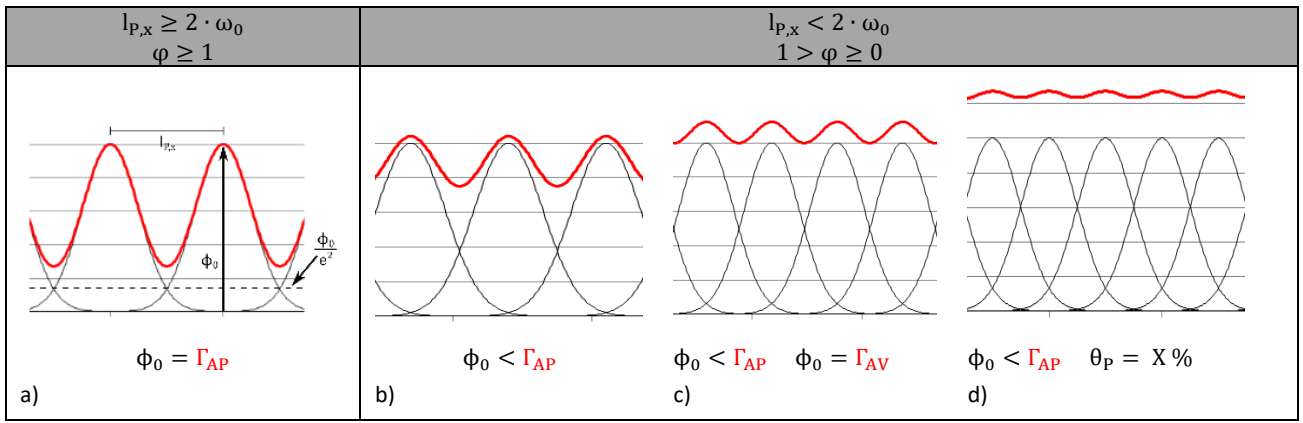

Fig. 4: Accumulated 1D fluence profiles (red) for line ablation with a) No pulse overlap and b) Pulse overlap such that peak accumulated fluence is practically still equal to maximum spot fluence c) Pulse overlap such that valley accumulated fluence equals maximum spot fluence d) Large pulse overlap with accumulated fluence profile can be considered "flat" and overshoot indicator is small e.g. below a certain threshold.

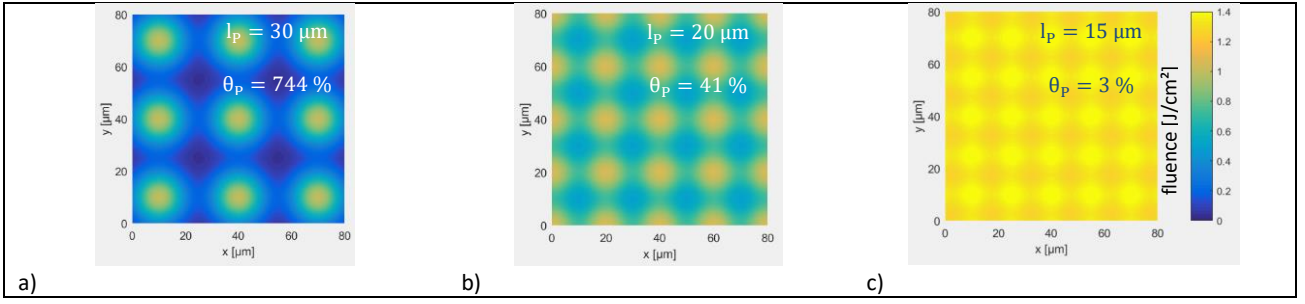

Fig. 5: 2D areal accumulated fluence distributions ($\phi_0$ = 1 J/cm², $\omega_0$ = 16 µm) with a) Large gaps between individual spots such that fluence level locally drops to almost zero b) overlapping pulses resulting in accumulation of pulse fluences s c) pulse spacing similar to the pulse radius resulting in a practically homogeneous (flat) accumulated fluence profile far above $\phi_0$.

## 2.2 The ablation pattern

Process parameters are considered well suited, if they allow complete homogenous ablation of a metallic thin film. At the same time, the underlying polyimide substrate must stay transparent indicating that surface damage is avoided. This is possible only if the accumulated fluence is homogeneously spread. To obtain a texture free homogeneous areal ablation, an isotropic and practically homogeneous 2D fluence distribution is required. The isotropic fluence distribution requirement excludes some of the ablation patterns that are in practical use. Here we use the #-pattern as shown in Fig. 2 where horizontal lines are vertically stacked at a pitch of $l_{P,y}$.

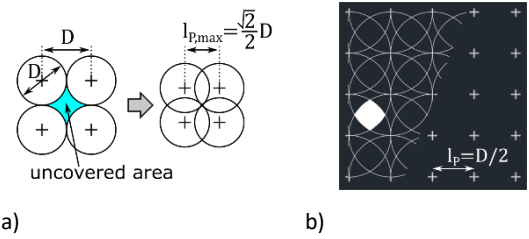

Fig. 6: Illustration of #-pattern a) explaining the maximal pulse distance to prevent uncovered areas of the specimen b) providing an example pattern where the borders of overlapping pulse areas and spot centers are marked. The white area is receiving fluence contributions from four single pulses.

As seen in Fig. 6a, the maximal pulse distance $l_{P,max}$ to completely cover an area with the "#-pattern" is given by $0{,}71 \cdot D$. With increasing pulse overlap ($\varphi \rightarrow 0$) more area on the substrate is irridiated multiple times. This makes the comparison of single pulse thresholds with ablation thresholds for line or areal ablation a challenging task as the incubation effect must be considered.

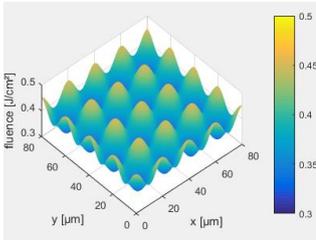

Fig. 7: 2D fluence distribution obtained with a "#-pattern" at $l_P = 20$ µm $= 0{,}63 \cdot D$. As this is smaller than $l_{P,max}$, all areas between pulses are irradiated multiple times.

## 2.3 Further relevant ablation parameters

Liu et al. already stated that any laser ablation mechanism starts with the absorption of laser energy in the target [5, p. 1707]. Because metals show high absorptivity, light is fully absorbed within nanometers below the surface [6], while typical coating thicknesses are beyond 50 nm. For a pure Nickel coating, penetration depth was found to be approximately 50 nm using Lambert-Beer's law [6], [13, p. 339]. The coating on our samples was fabricated with a sputter target made with 80 wt% Nickel and 20 wt% Chromium, such that the penetration depth of nickel gives a good approximation for the NiCr alloy.

Pulse radius as well as threshold fluence are important factors for laser induced ablation processes. Liu et al. have found a linear relation between the square of the optically measurable diameter of the laser modified zone D and the fluence plotted on a logarithmic scale [14, p. 197].

$$D^2 = 2 \cdot \omega_0^2 \ln\left(\frac{\phi_0}{\phi_{th}}\right) \quad (2\text{-}8)$$

This relation is commonly used to calculate (by squared error regression with $y = a \cdot \ln(x) + b$ [6]) effective pulse radius and threshold fluence for bulk material abrasion or laser-induced surface modifications but has some limitations when it comes to thin film ablation

Jee et al. have shown that material specific threshold fluence diminishes with increasing number of incident pulses [15]. This holds true for coating and substrate, which is why large pulse overlap ablations often result in substrate damage. The law of incubation can be stated as a formula, using the so called incubation factor $\xi$.

$$\phi_{th}(N_P) = \phi_{th}(1) \cdot N_P^{\xi-1} \quad (2\text{-}9)$$

A value of $\xi = 1$ means that no incubation influence can be observed. Typical values of $\xi$ for metals are between 0,8 and 0,96 [6],[1, p. 1160].

## 3 Results of single spot ablation

From theory, there are two ways to achieve homogeneous ablation. On the one hand, the substrate can be irridiated with low fluence and a high pulse overlap resulting in a smooth accumulated fluence profile. In this case, the accumulated fluence level is much higher than the maximum spot fluence value $\phi_0$. On the other hand, lower pulse overlap values allow for higher single pulse fluence levels (which showed to be beneficial) while the overall fluence variance ($\theta_P <$ e.g. 1 %) is still acceptably low.

We show that selective ablation results for line- and/or areal ablation can be achieved, if laser fluence and pulse overlap are adjusted properly. To describe the selective ablation criteria in case of thin film ablation the minimum level of the accumulated fluence profile $\Gamma_{AV}$ will be related to the previously determined accumulated single spot threshold fluence. As the threshold fluence is dependent on the number of incident pulses, the so called incubation factor must also be determined. Thus, the material specific threshold fluence $\phi_{th}$ and the incubation factor $\xi$ of the coating are identified at first. This is accomplished with the aid of multiple D²-experiments where the coating is irradiated with different numbers of single laser pulses $N_P$ at increasing energy. In order to gain the material specific incubation factor, the threshold values for different numbers of pulses are plotted such that a line fit can be conducted. For these experiments the laser energy is configured, such that a single shot $N_P = 1$ can fully trepan the coating at the highest fluence level used. We use confocal laser scanning microscopy to measure diameter and ablation depth at the irradiated spot. The energy per pulse is varied in the range of 0,29 µJ to 4,83 µJ, giving fluence values between 70 mJ/cm² and 1,11 J/cm².

### 3.1 Ablation zone modell

An ablation spot on a thin film coated polymeric substrate is not clearly confined and multiple characteristic zones can be observed. Fig. 8 shows three distinguishable ablation zones. Outer areas with low fluences only show surface modifications and cracks in the coating. We can define the size of this influence zone by the expansion of the cracks. Higher fluences yield a ring-shaped area of selective ablation in which the substrate is unveiled but not damaged. This will be called selective ablation zone. For the highest fluence around the spot center a destruction zone evolves, which is recognized optically by a milky and/or rough appearance of the polymeric substrate. Fluence thresholds for each of the three zones have to be determined.

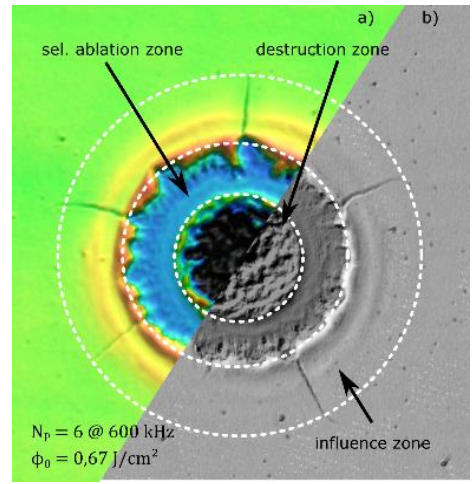

Fig. 8: LSM micrograph (x150) showing three distinguishable ablation zones for a NiCr-coated sample (100 nm) with two different imaging technics. a) 3D-image b) Differential interference contrast-image (DIC)

### 3.2 Single pulse experiment

The subsequent development of the ablation zones is illustrated by micrographs shown in Fig. 9. In Fig. 9a the fluence is small so that only some surface modification of the coating can be observed. With increasing fluence a clear influence zone with cracks starts to appear (Fig. 9b). Close to the selective ablation threshold fluence the coating lifts up and forms a bladder (Fig. 9c). The coating is sometimes completely removed in this area due to the particle suction system inside the laser machine. Further increase in fluence allows to generate a clear selective ablation zone as shown in Fig. 9c. The step between coating (green area) and substrate (light blue) is in the range of 100 nm which is the thin film thickness.

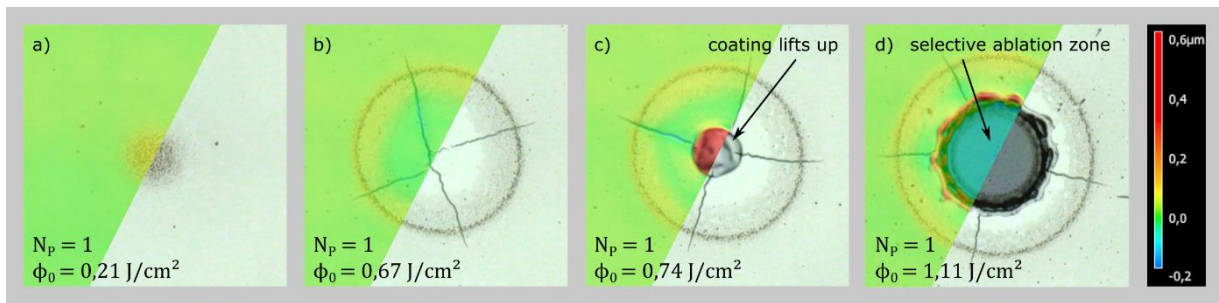

Fig. 9: Results of single pulse experiments with increasing fluence (left: 3D reconstructed image / right: direct optical view) a) First material reaction b) Cracks form inside the influence zone c) The coating lifts up d) Selective ablation showing no substrate damage

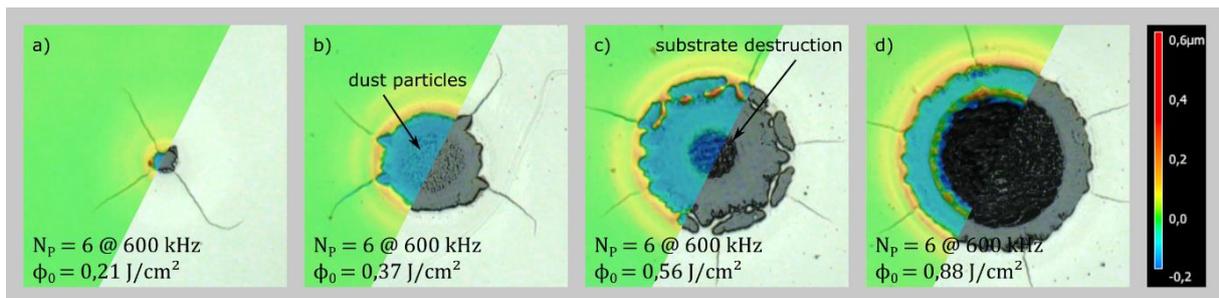

Fig. 10: Results of multiple (six) pulse experiments with increasing fluence (left: 3D reconstructed image / right: direct optical view) a) Large influence zone with cracks b) Selective ablation zone, dust particles remain but no substrate damage can be detected c) Substrate damage starts to appear d) Selective ablation zone becomes narrow

## 3.3 Multiple pulse experiments

Incubation experiments with multiple pulse repetitions (at φ=0) are carried out with gradually increasing pulse energy where the increasing size of the influence, the ablation and destruction zone are observed. For small fluences a cracked influence zone with a small selective ablation area becomes visible on the surface (Fig. 10a). Note that laser fluence is equal to the one used for the single pulse experiment in Fig. 9a. With increased fluence the diameter of the influence zone and selective thin film ablation zone in the spot center becomes larger (Fig. 10b). Dust particles from the coating sometimes remain on the substrate. These can be cleaned using an ultrasonic bath or similar methods. With even higher fluence substrate damage starts to appear (Fig. 10c). In Fig. 10d the destruction zone has expanded drastically such that the zone of selective areal ablation is narrowed.

The measured diameters of different experiments are used to draw the D²-plots in Fig. 11. The plots show only the data of the selective ablation zone. Selective threshold fluence for $N_P = 80$ is found to be $\phi_{th,80} = 0,07 \text{ J/cm}^2$.

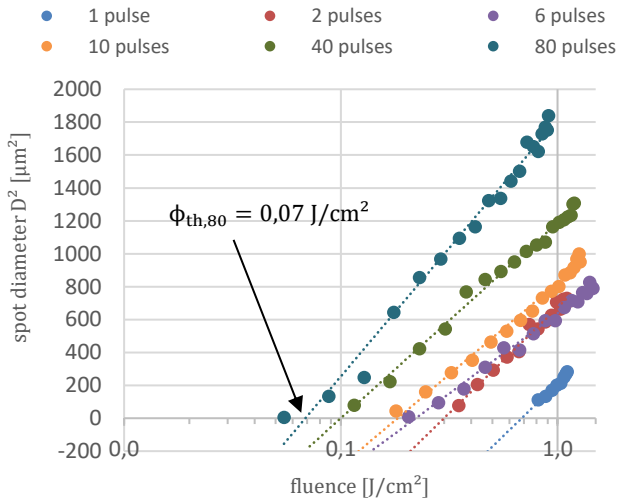

Fig. 11: D²-plot for a thin film selective ablation of 100 nm NiCr on polyimide

The accumulated fluence $\Gamma_{AP}$ that achieves selective ablation can be calculated for stationary spot experiments by multiplying the selective threshold fluence with the number of irradiated pulses.

$$\Gamma_{AP} = \phi_{th}(N_P) \cdot N_P \quad (3\text{-}1)$$

If number of irradiated pulses and the chosen fluence are well picked, full selective ablation with no substrate damage or dust generation can be achieved (see Fig. 12).

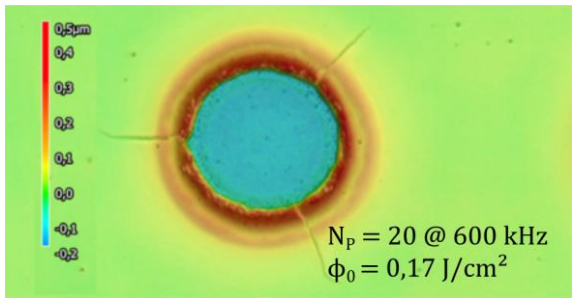

Fig. 12: 3D-image proving that full selective ablation is achievable

In section 2 we already calculated the gaussian spot size to be $\emptyset_{0,calc} = 40,1 \text{ µm}$. In comparison, the evaluation of the D²-experiment in the selective ablation regime leads to a spot radius of only $\omega_{0,D^2} \approx 16 \text{ µm} \Rightarrow \emptyset_{0,D^2} = 32 \text{ µm}$.

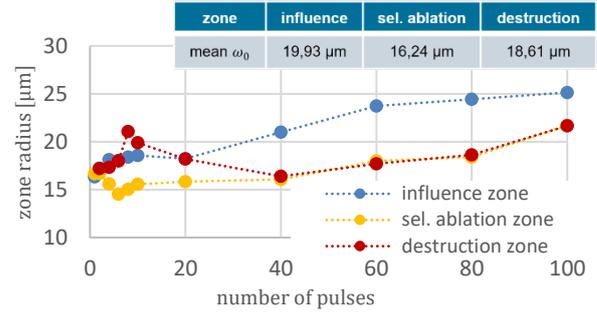

Fig. 13: Plot of different zone radii evaluated from D²-experiments

In Fig. 13, it can be seen that the D²-evaluation tends to yield higher values if the number of incident pulses increases. If plotted data is averaged for the three different zones, the radius of the influence zone $\omega_{0,D^2} \approx 20 \text{ µm} \Rightarrow \emptyset_{0,D^2} = 40 \text{ µm}$ correlates best with the calculated spot size. This means that influence zone data is most valuable to define the spot size, whereas selective ablation zone data must be used to define threshold fluences.

## 3.4 The incubation effect

From our D²-experiments of NiCr we find the incubation factor for selective ablation to be $\xi_{NiCr, Abl} = 0,53$. This is smaller than typical values for metallic thin films (see section 2.3). So we conclude that selective ablation is mainly driven by interface destruction between coating and substrate. Fig. 14 summarizes the threshold fluences for the three ablation zones defined in section 3.1 at increasing pulse numbers. It can be seen, that the threshold fluence of ablation and destruction zone are identical for number of pulses ≥ 40. Even though selective ablation can also be achieved in a narrow window at higher pulse numbers, we interpret this as a suitable upper boundary condition.

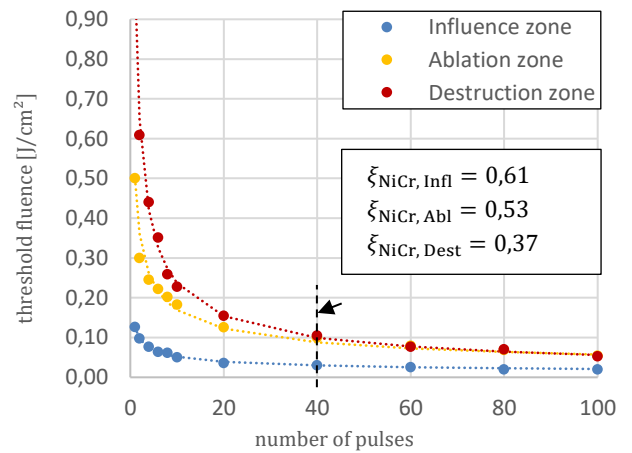

Fig. 14: Plot of threshold fluences for different ablation zones with increasing number of pulses at 600 kHz. Incubation factor $\xi$ can be derived from these curves.

Laser frequency, mark speed and pulse divider must be chosen in such a way, that the number of pulses hitting the very same position on the specimen stays below the upper boundary condition ($N_P \geq 40$ in this case). Otherwise substrate damage is likely to occur.

## 3.5 Influence of laser frequency

The laser pulse frequency is not considered to influence the ablation threshold. Our experiments however show that ablation behavior changes for higher numbers of pulses depending on laser frequency (see Fig. 16). This demands for an adjustment of the current well known methodology for ablation threshold determination developed by Liu et al. [2] which is not considering frequency influences.

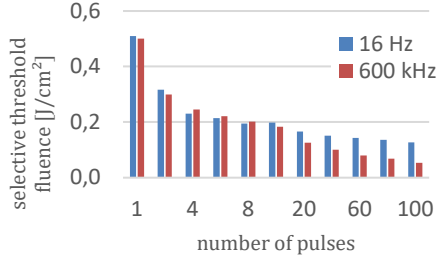

Fig. 15: Direct comparison of D²-experimental data confirms that ablation thresholds are influenced by laser frequency only at higher number of pulses ≥ 20.

In Fig. 15, it can be seen, that the selective threshold fluence for smaller number of pulses (1-10) is indeed frequency independent. For higher pulse numbers (≥20) however, threshold values obtained using equation ( 2-8 ) for 16 Hz and 600 kHz differ. An incubation factor $\xi_{\text{NiCr, Abl,16Hz}} = 0{,}67$ for the selective ablation zone is obtained at 16 Hz pulse frequency. Looking at Fig. 16 this phenomenon shows. As the threshold calculation is based on the measurable ablated zone diameter, the values gained from the two upper rows of the figure are very similar. In contrast, in case of higher pulse numbers (the two lower rows) the influenced zone is always larger for the higher pulse frequency. Also the selective ablation and destruction zone have joined and cannot be perceived separately. Thus, the purely selective ablation is absent at higher fluences. The transition from purely selective ablation to substrate damage becomes more sudden at higher pulses numbers.

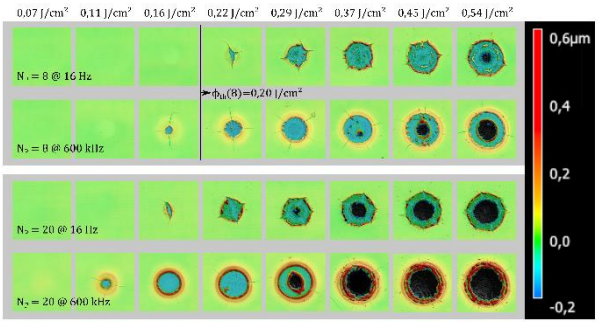

Fig. 16: Comparison of multiple pulse experiments at increasing single pulse fluence. In case of $N_P = 8$ (two upper rows) the ablated diameter is very similar and independent of laser frequency. At the bottom rows $N_P = 20$, the selective zone ablated at 600 kHz is always larger than at 16 Hz. Black areas indicate severe substrate damage where the ablation crater is deeper than 200 nm.

In addition, optically verifiable ablation happens at lower fluence when pulse frequency is high, even though the calculated values match. We assume this to be the results of different delamination effects. In case of low frequency ablation, the coating is "slowly" pulverized leading to an increased amount of residual particles and a poorly definable ablation zone border. For high frequency ablation the coating experiences a flat lift and is removed in one pieces or larger flakes. Thus at the boundary of the ablated zone the coating bulges upwards.

## 4 Selective ablation in scan patterns

In the past suitable material ablation process conditions were typically found after a fixed scan strategy was defined. After that, an ablation threshold value is experimentally obtained for ablation with the chosen scan strategy. We will now discuss a more systematic approach to define perfect conditions for selective ablation of thin metal films on susceptible polymeric substrates based on experimentally obtained single pulse thresholds. To allow for good transfer from single pulse threshold to line or areal pattern ablation only scan patterns with practically isotropic and homogeneous fluence distribution are considered. In the following, we will derive rules on the basis of single pulse thresholds and show how they apply for the exemplary case of thin film nickel-chromium ablation. We do not discuss multiple repetitions of the same ablation pattern as they do not improve the surface quality obtained after ablation. They are only necessary in cases of thicker layers to be removed.

### 4.1 Line ablation

Single laser-lines structured into a metallic thin film coating can be used to electrically insulate certain regions of the coating. This can yield sensor structures or help to separate conductive areas. Because very little material must be removed, line ablation, which we call contour cut [16, p. 82] is a time and cost effective technique. For line ablation laser pulses are placed along a line – each pulse overlapping its predecessor as defined by the parameter φ.

In line ablation, the average number of pulses $\widetilde{N}_P$ that hit a single position on the specimen can be defined as the product of pulse frequency f and the spot diameter $\emptyset_0$ divided by the mark speed v and pulse divider value PD. The pulse divider periodically guides a certain amount of pulses into a beam dump. The pulse frequency which is effective on the substrate can thereby be changed without changing the pulse generation frequency (600 kHz for our machine) and the single pulse fluence. PD=1 means that all pulses pass through the device such that they leave the source at their generation frequency.

$$\widetilde{N}_P = \left\lceil \frac{f \cdot \emptyset_0}{v \cdot \text{PD}} \right\rceil \qquad ( 4\text{-}1 )$$

Fig. 17a shows a micrograph with transmitted light illumination of a large line pattern test grid. Single pulse fluence is increased from left to right while mark speed is increased from the bottom to the top of the grid. Each "box" (the boxframe is solely a position marker) contains two scan lines having the same pulse spacing. The upper line is written at a pulse frequency of 600 kHz. The lower line consists of pulses placed on the specimen at a lower frequency of only 16 Hz. The pulse fluence and average number of pulses $\widetilde{N}_P$ is equal for both lines. In Fig. 17b the experimental results of the optically measured cut width are compared with the calculable theoretical values.

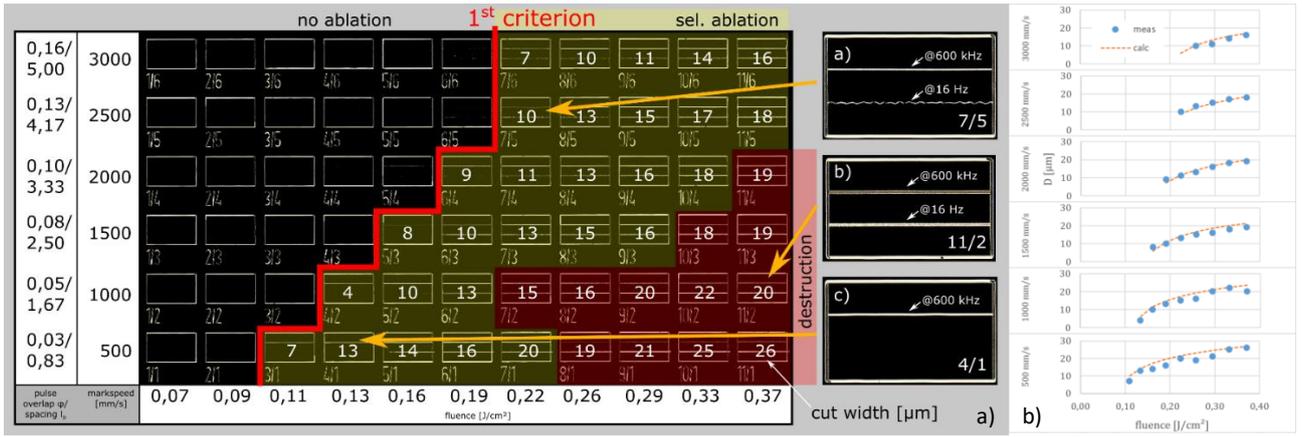

Fig. 17: a) Line ablation grid. Inside the box frame, there are two ablated lines: Upper line is marked at 600 kHz, lower line is marked by placing individual pulses in a row yielding a slow frequency of 16 Hz (pulse fluence $\phi_0$ and the number of pulses on a single spot $\widetilde{N}_P$ are the same for both lines). White numbers indicate cut width of lines [μm] ablated at 600 kHz. Side pictures a) A fluence of 0,22 J/cm² and a mark speed of 2500 mm/s allow for a selective line ablation using the standard 600 kHz provided by the laser source. In contrast, the line ablated at 16 Hz is intermittent b) The upper line shows severe substrate damage while the lower line is still in the selective region. Edgeline quality of lower line is low c) At these parameters the upper line is selectively ablated whereas no ablation is observed in the case of slow pulse frequency. b) Comparison of cut width D calculated for mark speed v at increasing fluence $\phi_0$ using equations ( 2-8 ),( 2-9 )and ( 4-1 ) with cut width measured optically.

Three different areas are separated (indicated by no, yellow and red color). The area on the left (no color) corresponds to conditions not leading to any ablation. The middle region (yellow) is the selective ablation area. White numbers inside the fields represent the measured cut-width of the upper line in μm. The red zone on the right marks those fields, where substrate damage is detectable by optical microscopy. Only the lines written at 600 kHz determine to which area (no color, yellow or red) the corresponding parameter set of the box is assigned to. From the grid the single pulse fluence values $\phi_0$ at which selective ablation starts to occur with respect to the mark speed (e.g. $\phi_0 = 0{,}16 \frac{J}{cm^2}$ for a mark speed of $1500 \frac{mm}{s}$) can be read off. The border between either no or selective ablation is given by the following criterion (red line in the plot).

Selective ablation criterion for line-ablation:

$$1^{st} \text{ criterion:} \quad \phi_0 \geq \phi_{th}(\widetilde{N}_P)$$

The single pulse ablation thresholds (see Fig. 15) show up to $N_P = 10$ overlapping pulses no dependency on pulse frequency. However we see a pulse frequency dependence in our line patterns (the two lines in each box can appear differently) which cannot be explained without assuming a difference in delamination effects, which likely have to do with progressing interfacial damage. Fast successive pulses still beyond the threshold for selective ablation already cause a delamination and result in flaking of the coating. However, this effect does not occur with low pulse frequencies. Nevertheless, this is important to note as this interferes with any attempt to forecast ablation behavior at very low frequencies. In our data we notice this frequency dependency only, if pulse overlap was quiet large (as for the grid in Fig. 17). For reduced pulse overlap ablation results for high and low pulse frequencies converge.

The graph in Fig. 18 contains the threshold fluence values for selective ablation and substrate destruction determined in the previous D²-experiments (stationary spot) in dependence of the number of incident pulses. They define the window for selective ablation with $\widetilde{N}_P$ as the decisive parameter in line pattern ablation.

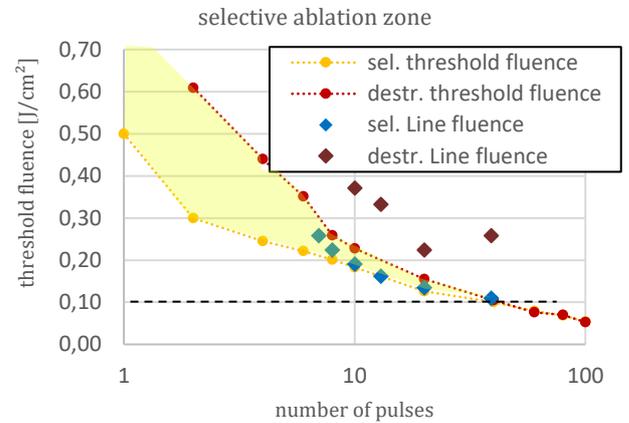

Fig. 18: Logarithmic plot of fluence threshold values as obtained by single spot (multiple pulses) which define the selective ablation window (yellow) for selective line ablation.

The blue diamonds mark the fluence level $\phi_0$ in the plot where visible ablation starts to occur in dependence of averaged number of overlapping pulses $\widetilde{N}_P$. All these data points lie inside the proposed selective ablation zone. The knowledge of the pulse fluence $\phi_0$, the spot size and the threshold fluence corrected by the incubation effect allow the calculation of the resulting cut width D (see Fig. 1). The plot in Fig. 18 shows that experimental data and ablation theory align.

## 4.2 Areal ablation using #-pattern

In order to predict areal ablation results the incubation effect plays an important role because the 2D-irradiation pattern causes exponentially more pulse overlaps than for line ablation. Depending on the ablation pattern (here we focus on the #-pattern as described in paragraph 2), the number of laser pulse hits on a single position must be determined. This value can be calculated knowing the pulse spacings used in an experiment. In order to do so, for each laser pulse of the ablation pattern, a cylinder on the sample surface with a height of one is modeled.

By adding up the heights of all distributed cylinders on the surface, the number of laser impacts at any position Z(x, y) can be determined as:

$$\Delta Z = \begin{cases} 0 \text{ if } \sqrt{(x - l_P \cdot n_x)^2 + (y - l_P \cdot n_y)^2} \geq \omega_0 \\ 1 \text{ if } \sqrt{(x - l_P \cdot n_x)^2 + (y - l_P \cdot n_y)^2} < \omega_0 \end{cases} \quad (4\text{-}2)$$

$$Z(x,y) = \sum_{n_x} \sum_{n_y} \Delta Z(x, y, n_x, n_y)$$

$$n_x \in [0, \dots, N_x], n_y \in [0, \dots, N_y]$$

A Z distribution calculation is exemplified in Fig. 19. Meanvalue of incident pulses is two for the shown parameter set.

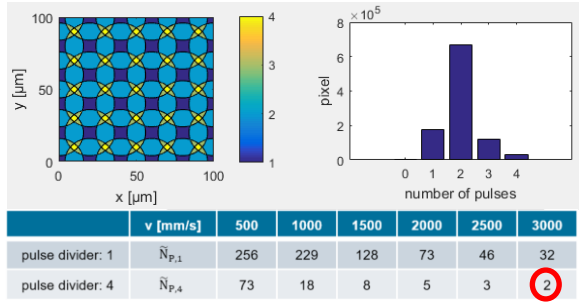

| v [mm/s] | 500 | 1000 | 1500 | 2000 | 2500 | 3000 |
|---|---|---|---|---|---|---|
| pulse divider: 1   $\widetilde{N}_{P,1}$ | 256 | 229 | 128 | 73 | 46 | 32 |
| pulse divider: 4   $\widetilde{N}_{P,4}$ | 73 | 18 | 8 | 5 | 3 | 2 |

Fig. 19: Map and histogramm of the specimen. In this case, a #-pattern with $\omega_0$=16 µm and a pulse spacing of $l_P$=20 µm (results from a mark speed of 3000 mm/s, a laser frequency of 600 kHz and a pulse divider of 4 is evaluated. Those regions which are hit by most pulses are marked yellow, whereas those areas struck only once remain blue. The histogram on the right shows how large is the surface fraction (in arbitrary units) that is hit by a certain number of laser pulses

From a histogram as shown in Fig. 19 the average number of pulse hits $\widetilde{N}_P$ on the surface can be calculated (weighing them with their corresponding area). Depending on the pulse overlap, the mean number of pulses per spot always differs assuming a constant spot diameter.

We create an areal ablation parameter grid to test ablation thresholds determined by single spot experiments with practical data from areal ablation. Fig. 20 shows a parameter grid for areal ablation using the #-pattern. Fluence and mark speed are adjusted in the same manner as for line ablation (look back to Fig. 17). It can be seen, that areal ablation with a single pulse fluence of $\phi_0 = 0{,}11$ J/cm² causes interface destruction between coating and underlying substrate (field 3/3). This ablation mechanism is denoted "flaking". The polymer substrate appears very clear and bright after flaking which indicates a damage free surface. Even though selective areal ablation in this regime seems promising, it has to be taken into account that for geometric shapes more complicated than squares we observed flakes to adhere much stronger.

Right of the flaking region is the ablation zone appears. Laser parameters for field 9/5 have yielded a complete coating removal, even though unveiled substrate is not as bright as for field 3/3 (flaking ablation). Laser parameters of field 11/5 cause a distinct substrate damage that can be seen by the color change. Any substrate damage comes along with surface roughness scattering the through light. Thus, the darker the substrate appears the more severe is the damage. Ablation by flaking can provide ideally preserved substrate surfaces but appears to be inappropriate for many applications as flake residues can cause electrical short circuits. In the ablation area to the right of the flaking regime, however, the substrate appears comparatively dark. The explanation can be found in Fig. 14, where differences in the threshold fluences for selective ablation and damage almost vanish when the mean number of incident pulses is equal or greater than 40.

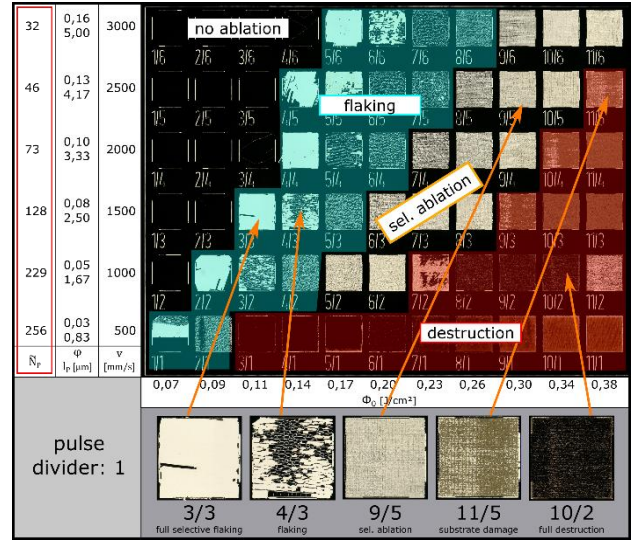

Fig. 20: Areal ablation grid with pulse divider 1 at 600 kHz pulse frequency. Fluence is increased left to right. Values on the left indicate the mark speed dependent mean number of pulses $\widetilde{N}_P$ for each row. Blue area marks all those fields where the coating has lifted in flakes from the substrate. All fields inside the selective ablation zone show some unwanted greyish substrate color. This is because of residues and dust. Nevertheless, they are electrically isolating. The red area defines those fields where substrate damage has occurred. The five micrographs on the bottom enlarge some of the fields inside the grid.

A simple way to reduce the number of pulses is the use of a pulse divider. With this, the pulse overlap can be decreased without affecting the fluence. Fig. 21 shows the corresponding parameter grid with apulse divider of four. As the level of the accumulated fluence profile is much lower as compared to the case of pulse divider one (in Fig. 20) the single spot fluence can be much larger without causing substrate defects (therefor the grid is extended to the right). In this case five regions can be identified. Right of the flaking zone appears a dotting zone (green area). "Dotting" means, that selective ablation is achieved only in the center of the pulses, but pulse spacing is such, that unveiled circles do not intersect and coating residues remain between them.. There is a smooth transition from the dotting zone to the selective ablation zone. Laser parameters in the selective ablation area allow flake free ablation of coatings without damaging the substrate. If one compares the substrate coloration in the flaking region (4/1) with the ablation results in the selective region (e.g. 11/2, 4/3 and 7/4), there is hardly any difference. No selective ablation results can be detected in the bottom line of the grid because the mean number of pulses (in this case $N_P = 73$) is too large, such that selective ablation threshold of the coating and destruction threshold of the substrate overlap in the incubation plot (see Fig. 14). In conclusion, flaking and destruction region are direct neighbors.

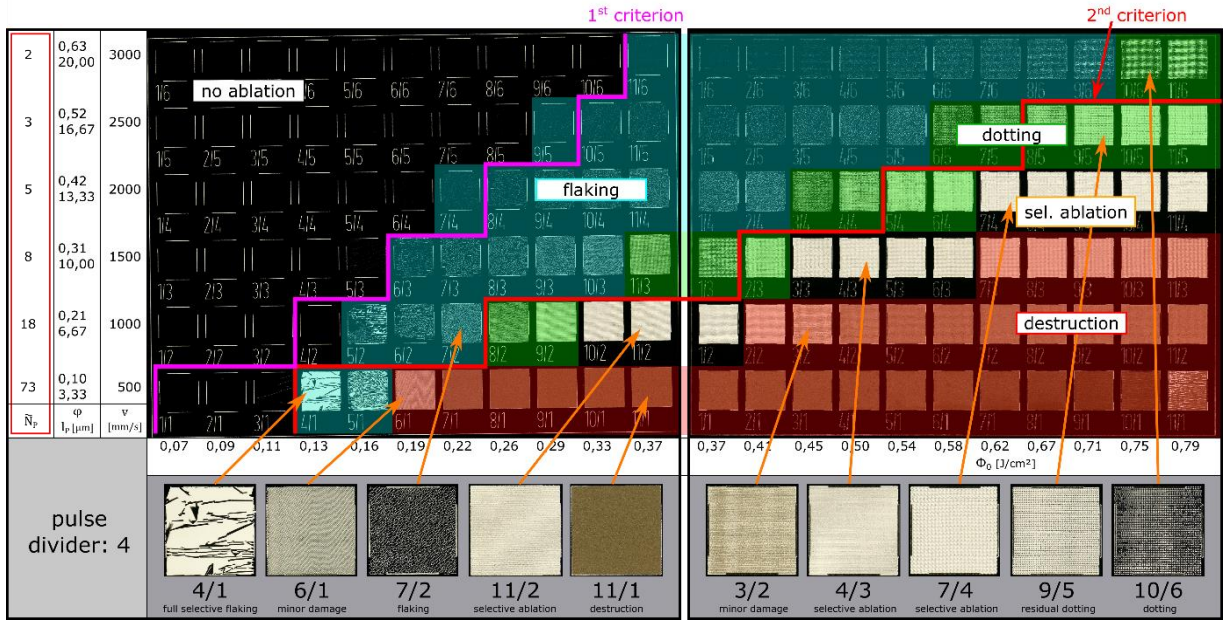

Fig. 21: Areal ablation grid with pulse divider 4 at 600 kHz pulse frequency. Fluence is increased left to right. Mean number of pulses $\widetilde{N}_P$, pulse spacing, pulse overlap and mark speed are indicated on the left. Five ablation zones can be differentiated: no ablation, flaking, dotting, selective ablation and destruction regime. In the dotting regime the coating shows a uniform pattern of holes but retains areal conductivity. The two criterion boarders are marked inside the grid. The ten micrographs at the bottom enlarge fields inside the grid which are of particular interest.

The first criterion, validated during the evaluation of the line experiments, compares the calculated fluences in the pattern with the pulse number dependent selective ablation threshold obtained in stationary beam experiments. The boundary given by this criterion is now also indicated in the parameter grid for areal patterning in the form of a pink boundary line. However, it does not allow an exact delimitation of the selective ablation range. One reason for this deviation is a simplification in the calculation of $\widetilde{N}_P$. The Gaussian profile of the laser is disregarded so far. Therefore, the accumulated valley fluence $\Gamma_{AV}$ as defined in section 2.1 is considered in the following to account for the Gaussian beam profile.

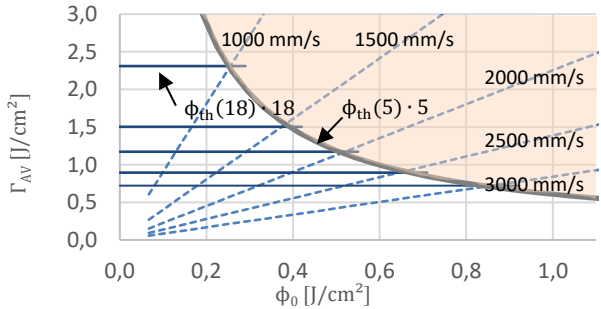

Fig. 22: Accumulated fluence plot. Yellow shaded area marks the selective ablation zone. So if mark speed dependent accumulated fluence is larger than the product of ablation threshold and pulse number selective ablation is achieved.

Fig. 22 shows a plot where the product of ablation threshold and the number of pulses is related to the accumulated fluence profile. The ablation threshold $\phi_{th}(\widetilde{N}_P)$ defines the minimal fluence which is needed to achieve selective ablation with a number of pulses $\widetilde{N}_P$. Thus the product of experimentally obtained selective ablation threshold and mean number of pulses is equal to the minimal accumulated fluence level needed to achieve ablation. This value can therefore be compared to the calculated accumulated valley fluence $\Gamma_{AV}$. Only if the valley fluence is larger than the product of threshold times pulse number, ablation will occur everywhere.

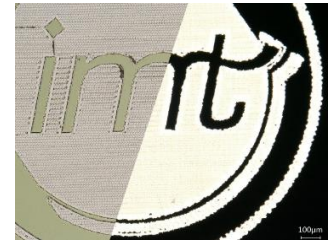

Fig. 23: Micrograph of successful areal ablation. Left side shows the sample directly after laser irradiation. Right hand side picture was taken after ultrasonic cleaning with through light illumination only. Coating dust is completely removed.

This second criterion is additionally indicated by a red line in the parameter grid. It exactly locates at the transition between dotting and selective ablation. In conclusion, we have experimental evidence to propose two simple criteria to predict ablation behavior in selective metal film removal from polymeric substrates.

To sum up, it can be said that the well-known first criterion which compares the laser fluence with the pulse number dependent ablation threshold is valuable to forecast the ablation results of single spot and pure line ablation. In case of areal ablation this criterion is not sharp enough though. To predict at which laser fluence level areal selective ablation can be achieved we formulate a second, sharper criterion as:

$$2^{nd} \text{ criterion:} \quad \frac{\Gamma_{AV}}{\widetilde{N}_P} \geq \phi_{th}(\widetilde{N}_P)$$

The described method for areal selective ablation can handle even relatively large pulse distances which may be necessary to keep the accumulated fluence at a level not too far above of what is needed for sufficiently homogeneous selective ablation. However, a too small pulse overlap can lead to noticeably rippled edges (as can also be seen in Fig. 23). A subsequent smoothing of the contour with low fluence but high pulse overlap can improve the edgeline quality.

# 5  Conclusion

A method has been established to provide laser parameters for full selective ablation of a metallic thin film (NiCr) on a susceptible polymeric substrate (polyimide). In particular, pulse overlap and pulse fluence have to be carefully controlled. This methodology based on experimentally derived ablation thresholds may help to efficiently find a working set of parameters aiming to minimize overall energy input into the substrate. This may greatly reduce the risk of substrate damage. In practice, it may be sufficient to calculate $\widetilde{N}_P$ for simple line ablations and work with the incubation dependent ablation threshold. In case of areal ablation, the second criteria must be considered. This allows to predict suitable parameters for an areal selective ablation.

1. Adjustment of laser setup (with $N_P = 1$, $\phi_{0,\max}$ such that ablation can be achieved)
2. Maximized laser frequency (this yields small pulse energy suited for thin films)
3. Multiple D²-experiments to define $\phi_{th}$ and $\xi_{abl}$
4. Calculation of $\widetilde{N}_P$ for line- or areal ablation
5. Comparison with ablation criterions
6. In case no appropriate set of parameters can be found: Change pulse divider to enlarge pulse spacing

**Acknowledgements**

Funding: This work was funded by the German Research Foundation (DFG) [grant number: DI 1934/8-1].